\documentstyle[prl,aps]{revtex}
\twocolumn
\begin{document}
\input{epsf}
\draft
\twocolumn[\hsize\textwidth\columnwidth\hsize\csname@twocolumnfalse\endcsname
\title{Breakdown of Scale Invariance in the 
Phase Ordering of Fractal Clusters}
\author{Massimo Conti}
\address{Dipartimento di Matematica e Fisica, Universit\'{a} di Camerino,
and Istituto Nazionale di Fisica della Materia, 62032,
Camerino, Italy}
\author{Baruch Meerson}
\address{The Racah Institute  of  Physics, Hebrew
University   of  Jerusalem,
Jerusalem 91904, Israel}
\author{Pavel V. Sasorov}
\address{Institute of Theoretical and Experimental 
Physics, Moscow 117259, Russia}
\maketitle
\begin{abstract}
Numerical
simulations with the 
Cahn-Hilliard equation show that coarsening of  
fractal clusters (FCs)
is 
not a scale-invariant process. 
On the other hand, a typical coarsening length scale
and interfacial area of the FC 
exhibit power laws in time, while
the mass fractal dimension remains invariant. 
The initial
value of the lower cutoff is a
relevant length scale. 
A sharp-interface
model is formulated
that can follow the
whole dynamics of a diffusion controlled growth, coarsening,
fragmentation and approach to equilibrium
in a system with conserved order parameter.
\end{abstract}
\pacs{PACS numbers: 61.43.Hv, 64.60.Ak, 05.70.Fh}
\vskip1pc]
\narrowtext

Non-equilibrium driven dissipative 
systems relax to equilibrium after
the driving agent is ``switched off" or depleted. 
In complex systems the relaxation dynamics 
can  be quite complicated, and it is natural to seek for
dynamic scaling and universality.  An 
instructive, exactly solvable nonlinear
example of dynamic scaling in
relaxation (coarsening) of rough (self-affine fractal) surfaces with 
{\it nonconservative}
dynamics is given by the deterministic
(undriven) KPZ-equation \cite{KPZ,BS}. A much older example 
is
decay of homogeneous and isotropic hydrodynamic 
turbulence \cite{Bat,Bar}. Finally, there is an important class of relaxation 
problems 
related to 
phase ordering dynamics, non-conserved and conserved,
in the bulk and on the 
surface \cite{Gunton,Bray,Zinke}. 

If the system obeys a conservation law, ``switching off"
of the driving agent occurs naturally. There are many important 
non-equilibrium systems 
that exhibit morphological 
instabilities and ramified growth at an early stage of the dynamics,
show phase ordering at an intermediate stage,
and finally approach a simple
equilibrium. A canonical 
example is provided by diffusion controlled
systems, such as deposition of solute from a supersaturated solution
and solidification from an overcooled
liquid. The stage
of morphological instability and its implications have been under extensive
investigation \cite{Langer,Kessler,Brener,Mineev,BMT}. If some noise is 
present, a fractal cluster (FC) can develop at 
this stage \cite{BMT}. The 
subsequent surface tension driven
coarsening of this FC is unavoidable in a closed geometry with a finite
amount of mass or heat. This stage
has not received much
attention, with the exception of the paper 
by Irisawa {\it et al}  \cite{Irisawa95} where  two-dimensional
Monte-Carlo simulations were performed, and
a power law found for the perimeter of a DLA
cluster versus time.

We are aware of two additional physical systems with a conservation law, for
which numerical simulations showed
nontrivial
fractal coarsening dynamics and power laws for the cluster perimeter: 
interface controlled \cite{Irisawa95,Irisawa96} and
surface diffusion controlled \cite{Irisawa95,Olivi96} systems. 
Besides, Stokes flow controlled coarsening has been discussed 
in the context of sintering of fractal matter \onlinecite{Olivi96}. 

No theory is available for any of these fractal coarsening 
systems, except for the very late post-fragmentation 
stage \cite{fragmentation}.  On the 
other hand, a FC is a particular case of 
disordered media
with long-range (power-law) spatial correlations \cite{Mandelbrot}. The
scaling hypothesis (SH) (the cornerstone of modern theory of phase
ordering \cite{scalinghyp}) does 
not exclude FCs when dealing with long-range 
correlations in the initial condition \cite{Bray}. Therefore, 
one is tempted to employ the SH and 
calculate the growth exponents for
the coarsening of FCs. We start with these simple calculations.
Then we
report our 
simulations of the diffusion controlled coarsening of a DLA aggregate,
as described
by the Cahn-Hilliard (CH) equation. Having measured, for the first time, 
the dynamics of the 
pair correlation function (which is very close to the 
average mass density) of the FC, we show
that the 
SH is invalid. On the other hand, we find that
a characteristic coarsening length scale and interfacial area of the FC 
exhibit power law dynamics (with a {\it new} growth exponent), while
the fractal dimension remains 
invariant 
(on an interval of scales shrinking with 
time). The initial
value of the lower cutoff of the FC is shown to be 
an additional relevant length scale. Finally, a 
minimalistic sharp-interface model is presented
that can follow the
whole dynamics of
the diffusion controlled system: an unstable growth,
coarsening, fragmentation and approach to equilibrium.

Let the initial state of a conserved 
system represent a single-connected, statistically
homogeneous 
self-similar mass fractal of the minority phase,
characterized by the fractal dimension 
$D$ on an interval of scales between the lower cutoff $l_0$ and upper
cutoff $L$. We 
start with a simple coarsening scenario \cite{Toyoki,Sempere} that is
{\it required}
by the SH.
It assumes that the fractal dimension of the
coarsening cluster remains constant on a (shrinking)
interval of scales between the time-dependent lower and upper cutoffs, 
$l(t)$ and  $L(t)$. The interfacial 
area $A$ and total mass $M$
of the FC are estimated as \cite{Mandelbrot}
\begin{equation}
A \sim l^{d-1} \, (L/l)^{D} \quad \mbox{and} \quad
M \sim l^{d} \, (L/l)^{D}\,,
\label{f1}
\end{equation}
respectively, where $d$ is the embedding Euclidean dimension.  
Mass conservation yields 
$L \sim l^{(D-d)/D}\,$ \cite{Sempere}.
Now assume that
$l(t) \sim t^{1/z}$.
Then we find the following 
scaling laws: $A(t) \sim t^{-1/z}$ 
and $L(t) \sim t^{(D-d)/D z}$. The scaling of $L (t)$
describes shrinking of the
FC in the process of coarsening \cite{Sempere}. 

Already at this stage a discrepancy appears: no shrinking has
been observed in any
direct numerical simulations of coarsening of FCs
\cite{Irisawa95,Irisawa96,Olivi96}. This 
gives a strong evidence for breakdown of 
scale invariance \cite{non-SI}. On the other hand, power laws 
for $A(t)$
reported in Refs. \onlinecite{Irisawa95,Irisawa96,Olivi96} 
indicate that the problem might possess scaling behavior 
of a more complicated nature. 

To clarify the matter,
we performed more detailed numerical simulations of a 
diffusion-controlled system. In addition to
$L(t)$ and $A(t)$, we 
followed the evolution of 
the equal-time pair correlation function (which is very close to
the average mass density of the FC, so we will not distinguish 
between them). Having measured it, one could
find the mass fractal
dimension and coarsening length scale of the FC for every moment of time.

If one remains, for one more moment, 
within the framework of the SH, one can easily predict the
dynamics of the mass density 
$\rho (r,t)$. At 
distances $r \ll l (t)$
from a (typical) reference point inside 
the cluster, the cluster is non-fractal: 
$\rho (r,t) \sim const$. At distances 
intermediate between $l$ and
$L$ 
$\rho (r,t) = a(t)\, r^{D-d}$,
where $a(t)$ is a function of time.   Matching these two asymptotics,
we have
$a(t) \, l^{D-d} = const$ and hence $a(t) \sim t^{(d-D)/z}$. Therefore,
for $l(t) \ll r \ll L(t)$ the SH predicts
$\rho(r,t) \sim (r/t^{1/z})^{D-d}$,
a simple self-similar expression. It is the absence of this
self-similarity that will enable us to 
utterly disprove the SH.

We concentrated
on the diffusion controlled coarsening and employed the CH
equation, a standard model of 
phase ordering with a conserved order parameter (COP) \cite{Gunton,Bray}:
\begin{equation}
\frac{\partial u}{\partial t} +
\frac{1}{2}\nabla^2 \left(\nabla^2 u + u - u^3 \right) = 0\,.
\label{f5}
\end{equation}
Eq. (\ref{f5}) was discretized and solved on the domain 
$0 \le x \le 512\,, 0 \le y \le 512$ with periodic boundary conditions.
We used an explicit Euler integration scheme to advance the solution
in time, and second order central differences to discretize the Laplace
operator. With a mesh size $\Delta x = \Delta y = 1$ no preferred directions 
emerged in the computational grid, due to the truncation errors;
a time step $\Delta t = 0.05$ was required for numerical stability. The 
accuracy was monitored by checking the mass conservation that was
verified in all the simulations within 0.01 \%.

We chose a DLA cluster \cite{Witten} as the initial condition. The 
fractal properties of DLA
clusters are somewhat more complex that those of a simple self-similar 
fractal  \cite{DLA}. However, it is a DLA-like
FC that can develop
during the 
diffusion controlled growth 
\cite{BMT}, so this choice is physically motivated. 
The initial clusters (like the one shown in 
Fig. 1a), with radius of order 250, were
prepared by a standard random-walk algorithm on a two-dimensional square
grid. To prevent fragmentation at an early stage of the
coarsening process, we followed 
the technique of Irisawa {\it et al.} \cite{Irisawa95,Irisawa96}: 
the aggregates were thickened by an addition of peripheral
sites. The mass fractal dimension was determined from 
the mass-radius relation \cite{Mandelbrot} 
and ranged from 1.67 to 1.72.
  
We identified
the cluster as the locus where $u ({\bf r}, t) \ge 0$. 
The coarsening process was followed up to a time $t=5,000$. 
Typical snapshots of the coarsening process are shown in Fig. 1.
One can see that smaller features of the FC
are ``consumed" by larger features, while 
the global structure
of the cluster is not affected. To characterize the coarsening process, 
the following
quantities were sampled and averaged over 10 initial 
configurations: (1) the gyration radius of the cluster, 
(2) the circularly averaged 
pair correlation function 
$g(r,t) = 
\langle \left[u(r^{\prime} , t)+1\right]\, 
\left[u (r^{\prime} + r, t)+1\right] \rangle$,
(3) the cluster perimeter $A_1 (t)$,
defined as the sum of  
$\mid\nabla \,u({\bf r}, t)\mid^2$ over the whole domain, and (4) the cluster 
perimeter $A_2 (t)$,
defined as the number of broken bonds between the aggregate sites.

The gyration radius of the FC has been found to remain
constant within possible logarithmic corrections. 
Evolution of $g(r,t)$ is shown in Fig. 2. One can see 
that coarsening affects only the smallest lengths, while
the intermediate-distance power-law part remains ``frozen". It is evident
that $g(r,t)$ does not acquire a
self-similar form, so there is no scale invariance. On the other hand, the
mass fractal dimension
remains invariant on an interval $l_c (t) \ll r\ll L$. The dynamics of the
coarsening length scale $l_c (t)$, extracted for 
each moment of time
from the slope of the linear part
of $g(r,t)$ (the Porod law \cite{Bray}),  are 
shown in Fig. 3. The late-time behavior
of the slope versus time shows a power law:  $t^{-\alpha_1}$
with $\alpha_1=0.19$. Therefore,
$l_c (t) \sim t^{\alpha_1}$, and the corresponding growth 
exponent $z_1 = 1/{\alpha_1}$
is close to 5 (and not to 3 as could be expected for a diffusion controlled
system with a COP \cite{Bray}). 
Fig. 4 shows the dynamics of the cluster perimeter estimates 
$A_1 (t)$ and $A_2 (t)$. The long-time 
dynamics of each of them is describable by a power law $t^{-\alpha_2}$, 
with $\alpha_2 = 0.19$ for
$A_1$, and $0.20$ for $A_2$. The corresponding result
of Monte-Carlo simulations \cite{Irisawa95} was slightly 
different: $0.22 - 0.24$. The close proximity of the exponents $\alpha_1$ 
and $\alpha_2$ gives evidence that it is a single exponent. 

\begin{figure}[h]
\vspace{0.0cm}
\hspace{0.0cm}
\rightline{ \epsfxsize = 9.0cm \epsffile{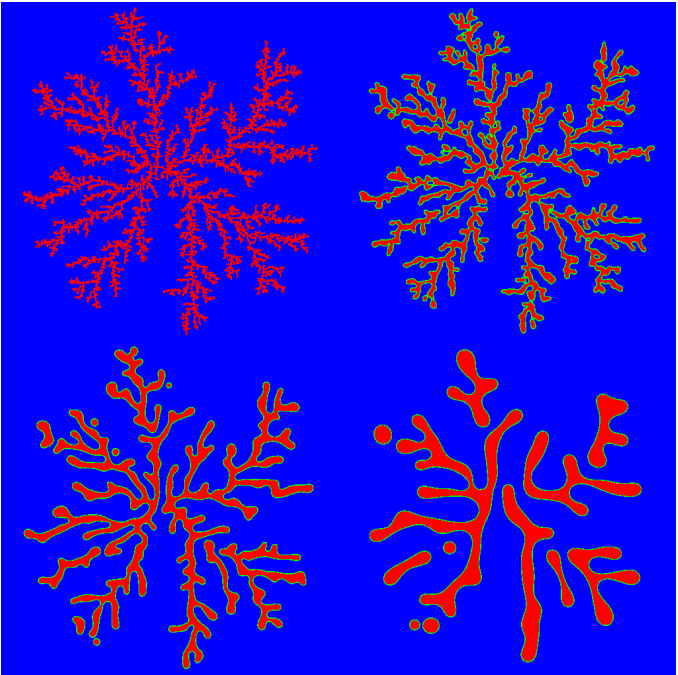}}
\caption{
Evolution of a DLA cluster undergoing 
coarsening in a conserved, diffusion controlled system. The upper
row corresponds to $t=0$ (left) and $34.7$ (right), the lower row to
$t=329.3$ (left) and $4,900$ (right).
\label{fig. 1}}
\end{figure}

\begin{figure}[h]
\vspace{-1.0cm}
\hspace{0.0cm}
\rightline{ \epsfxsize = 10.0cm \epsffile{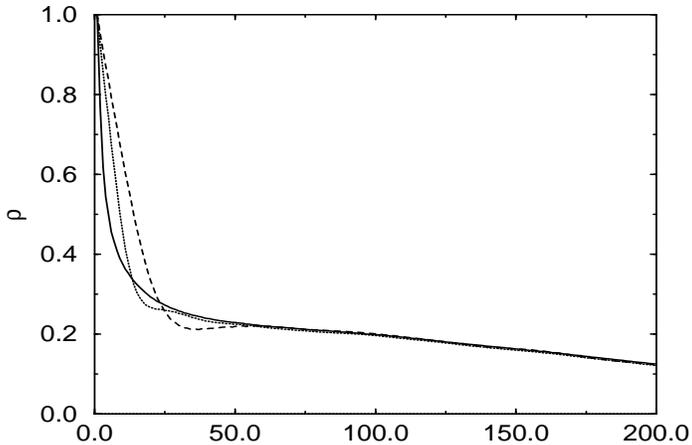}}
\caption{
Dynamics of the pair correlation 
function $g(r,t)$
for time moments $t=0$ (solid line),
$516.5$ (dotted line) and $4,900$ (dashed line).
\label{fig. 2}}
\end{figure}

Absence of scale invariance means presence of an additional length scale.
Function $g (r,t)$ gives evidence for the nature of 
this length scale. For our 
mass fractal at $t=0$ 
we have $g(r,\,t=0) 
\sim (r/l_0)^{D-d}$ in the fractal region 
$l_0 \ll r \ll L$. Preservation of the power-law 
part of $g$ with time (Fig. 2) implies that
the same asymptotics holds, on a shrinking interval
of radii, for $t>0$ (until fragmentation). That is, the
small intrinsic length scale $l_0$ remains relevant. How does it
show up in the phenomenology of coarsening?
Fig. 1 gives evidence that (i) the FC
can be
regarded as a set of ``bars", 
and (ii) the characteristic bar length $l_b$ 
grows in time faster than the bar width (identified with $l_c$).
The area of a single bar should scale like $l_b \, l_c$, hence the total area 
of the 
FC is
$l_b \, l_c\, (L/l_b)^D$. This quantity must be equal to the initial value of
the FC area, $l_0^2\, (L/l_0)^D$. This yields
$l_b \sim l_0 \,(l_c/l_0)^{1/(D-1)} \sim t^{\alpha_1/(D-1)}$. 

We will finish this Letter with formulating a sharp-interface
model that can describe the 
{\it whole} diffusion-controlled dynamics, from the stage of growth
through
coarsening  and fragmentation
to the final equilibrium. Consider a number of (possibly multiple-connected)
mass clusters characterized by a set of their (moving) interfaces $\gamma_i$.
Let now $u ({\bf r},t)$ be the mass concentration 
of the solution normalized to
the (constant) density of solute in the compact solid phase. 
The field $u$ in the liquid phase 
is governed 
by the diffusion equation 

\begin{figure}[h]
\vspace{-1.0cm}
\hspace{0.5cm}
\rightline{ \epsfxsize = 9.0cm \epsffile{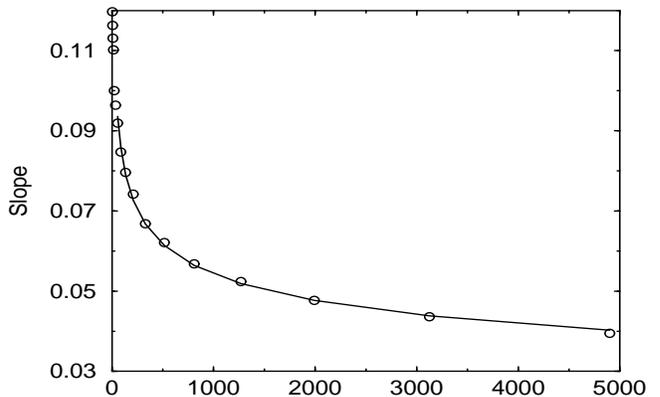}}
\caption{
The slope of the linear part of 
$g(r,t)$ versus time, and
its power-law regression.
\label{fig. 3}}
\end{figure}

\begin{figure}[h]
\vspace{-1.5cm}
\hspace{0.0cm}
\rightline{ \epsfxsize = 9.0cm \epsffile{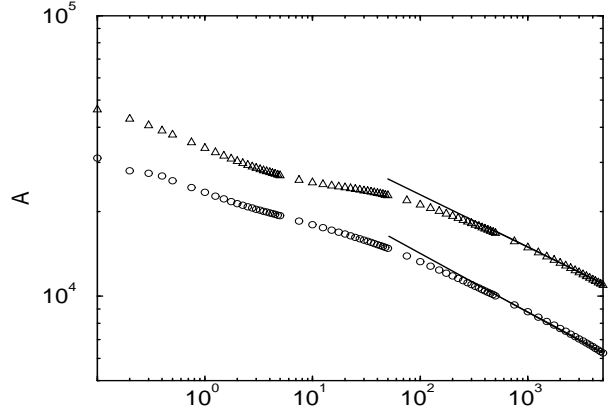}}
\caption{
Two estimates for the 
FC perimeter
versus time: $A_1$ (triangles) and $A_2$ (circles).
\label{fig. 4}}
\end{figure}

\begin{equation}
\frac{\partial u}{\partial t} = \chi \nabla^2 u
\label{a}
\end{equation}
in a finite $d$-dimensional domain.
We specify a no-flux boundary condition, $\nabla_n u \mid _{\Gamma} = 0$
on the external boundary $\Gamma$, where index $n$ stands for the normal 
component of a vector. Assuming that each of the 
interfaces $\gamma_i$ is in local
thermodynamic equilibrium, we employ the Gibbs-Thomson relation
$u\mid _{\gamma_i} = u_0 (1+\lambda_0 \kappa_i)$,
where  $u_0$ is the (normalized) equilibrium
concentration of the solution in the bulk, $\lambda_0$ is the 
capillary length and $\kappa_i$ 
is the local curvature
for $d=2$, or the mean curvature for $d>2$. (We limit 
ourselves to an isotropic surface tension.) 
Finally, mass conservation
at each of the moving interfaces 
yields the well-known relation for the normal speed: 
\begin{equation}
v_n^{(i)} = \frac{\chi \nabla_n u}{1-u}\mid_{\gamma_i}\,.
\label{1}
\end{equation}

It is easy to check that this model preserves the
total mass of the solute. In the normalized form  
\begin{equation}
\Omega_c + \int_{\Omega} u \, d {\bf r}  = \mbox{const} \,,
\label{2}
\end{equation}
where $\Omega_c$ is the total volume (area) of the
solid phase, while $\Omega$ denotes the region unoccupied by the solid
phase. This 
important conservation law 
does not appear in the more traditional theoretical formulations of the
diffusion-controlled growth 
problem \cite{Langer,Kessler,Brener,Mineev,BMT}, where an ``infinite" system
is studied, and the boundary condition 
corresponding to
a constant (positive) flux or constant supersaturation at
${\bf r}\rightarrow \infty$ is used. Notice that, even in the limit 
of strong diffusion,
it is
the 
full diffusion equation (rather than its Laplace's equation limit) and no-flux
condition on $\Gamma$ that provide the conservation law. Also, 
the usually small term $u$ in the denominator of 
Eq. (\ref{1}) should be kept to get Eq. (\ref{2}) right. 

In summary, we have demonstrated that diffusion controlled phase 
ordering of 
FCs is not a scale-invariant process.  
In spite of this, the problem possesses non-trivial scaling properties:
the coarsening length scale
and interfacial area of the FC 
exhibit power laws in time (with a new growth exponent), and 
the mass fractal dimension remains invariant. An 
additional small intrinsic length scale (the initial value of the
lower cutoff)
remains
relevant until the fragmentation stage. 
We believe that these findings apply to other coarsening
mechanisms as well. 

This work was supported in part 
by a grant from Israel Science Foundation, administered 
by the Israel Academy of Sciences and Humanities, and by the Russian Foundation
for Basic Research (grant No. 96-01-01876).



\begin{references}
\bibitem{non-SI} In the surface diffusion case, it 
has been noticed in Ref. \onlinecite{Olivi96} that
``... an intermediate shape cannot be approximated
as a self-similar fractal".
\end{references}
\end{document}